\documentclass[11pt]{article}

\newcommand{\bra}[1]{\langle #1|}
\newcommand{\ket}[1]{|#1\rangle}

\usepackage[margin=2.8cm, a4paper]{geometry}
\usepackage{enumitem}
\usepackage{dsfont}
\usepackage{amsmath,amsthm,amsfonts,amssymb}
\usepackage{fontenc}
\usepackage[all]{xy}
\usepackage{graphicx,subfigure,multirow}
\usepackage[figurename=Fig.]{caption}
\usepackage{tikz} 
\usetikzlibrary{shapes, arrows, shadows}
\usepackage[latin1]{inputenc}
\usepackage[scaled]{helvet}
\usepackage[toc, page]{appendix}
\usepackage{dsfont}
\newtheorem{thm}{Theorem}[] 
\newtheorem{define}[thm]{Definition}

\newtheorem*{new}{New Theorem}
\newtheorem*{conjecture}{New Conjecture}
\numberwithin{equation}{section}
\usepackage{cancel}

\usepackage[colorinlistoftodos]{todonotes}

\begin{document}

\begingroup 
\centering
{\Large\textbf{Computation in a general physical setting}  
\\[1.5em]
\normalsize  Ciar\'{a}n M. Gilligan-Lee\textsuperscript{$\ddagger$,}\footnote{Electronic address: ciaran.lee@ucl.ac.uk} }
\\[1em]
\begin{center}
\it  \it \textsuperscript{$\ddagger$} Department of Physics and Astronomy, University College London, UK. 
\end{center}
\endgroup

\begin{abstract}
\small{The computational abilities of theories within the generalised probabilistic theory framework has been the subject of much recent study. Such investigations aim to gain an understanding of the possible connections between physical principles and computation. Moreover, comparing and contrasting the computational properties of quantum theory with other operationally-sensible theories could shed light on the strengths and limitations of quantum computation. This paper reviews and extends some of these results, deriving new bounds on the computational ability of theories satisfying n-local tomography, and theories in which states are represented as generalised superpositions. It moreover provides a refined version of the conjecture that a quantum computer can simulate the computation in any theory within a certain sub-class of generalised probabilistic theories with at most polynomial overhead. The paper ends by describing an important relation between this conjecture and delegated computation, similar to the relation between quantum non-locality and device-independent cryptography.} 
\end{abstract}  

\section{Introduction}

Quantum theory is a strange beast; its predictions have been verified to unprecedented accuracy, yet the standard quantum formalism---in which quantum states are represented by positive semi-definite operators acting on an underlying complex Hilbert space---is as abstract as its predictions are accurate. Despite being universally accepted among physicists as a tool for calculating the probabilities of possible experimental outcomes, the standard language of complex Hilbert spaces and positive operators lacks direct physical or operational significance. As Asher Peres \cite{Peres} famously put it: ``Quantum phenomena do not occur in a Hilbert space. They occur in a laboratory''. 

Taking inspiration from Einstein's operational formulation of special relativity, in which reference frames are operationally defined using clocks and rods, researchers have begun to study quantum theory from an operational perspective. Researchers have formulated quantum theory within an operational framework of \emph{generalised probabilistic theories}---which generalise the probabilistic formalism of quantum theory. Remarkably, researchers have derived the structure of finite dimensional quantum theory \cite{hardy2005probability, hardy2011reformulating, chiribella2010probabilistic, chiribella2011informational, barrett2007information, barnum2017ruling} within this framework from operationally-framed physical principles---similar to Einstein's derivation of the Lorentz transformations from two physical principles: the constancy of the speed of light and the principle of relativity. Researchers have even used this framework to rule out certain a priori reasonable types of post-quantum physics \cite{lee2017no, lee2018beyond}.

A remarkable feature of this operational framework is that it provides examples of theories which differ from quantum theory, yet still make good operational sense. An obvious example is classical probability theory, which can be used to calculate probabilities for classicial situations such as tossing a coin, or conducting an experiemnt in the regime of Newtonian physics. More exotic examples\footnote{For further examples, see Ref.'s \cite{galley2016classification, Fermionic1, Fermionic2, Wootters-real, d'ariano2014determinism, massar2015hyperdense, janotta2011limits}.} include Spekkens' toy theory \cite{spekkens2007evidence, janotta2013generalized},  a construction colloquially known as ``Boxworld'' \cite{barrett2007information, popescu1998causality}, which achieves the largest possible violation of the CHSH inequality \cite{popescu-review} consistent with the no-signalling principle, and the theory referred to by the authors as ``Witworld'' \cite{cavalcanti2021witworld}, which exhibits post-quantum steering.    
 
The existence of such alternate theories allows for an investigation of the structural or information-theoretic properties of theories where different physical principles may hold. Such an investigation could provide a deep understanding of the possible connections between physical principles and information-theoretic advantages in a manner not wedded to the mathematical formulation of a specific theory \cite{chiribella2010probabilistic, barrett2007information, lee2016information, lee2016generalised}. This forms part of the broader research program of generalised probabilistic theories, which, in the words of Barnum, M\"{u}ller, and Ududec \cite{barnum2014higher}, aims to ``analyse the structure of physics---that is, the way that the different parts of physics fit together---by rigorously assessing the consequences of changing some of its parts''. Moreover, by comparing and contrasting the information-theoretic properties of quantum theory with other theories in this framework, one may shed light on the strengths and limitations of quantum information processing and quantum computing in particular.

The study of computation within the generalised probabilistic theory framework was initiated in the work of Barrett \cite{barrett2007information}. An intriguing aspect of that work resulted in a conjecture concerning the power of quantum computing relative to other theories within this operational framework. Specifically, it was conjectured that quantum theory may be able to simulate the computation in any generalised probabilistic theory with at most polynomial overhead. This conjecture has spurred much recent work \cite{lee2016generalised, lee2015computation, landscape, lee2015proofs, lee2016deriving, lee2016higher, lee2017oracles}. Ref.~\cite{lee2015computation} showed that in any theory satisfying the principle of \emph{tomographic locality}, which informally states that local measurements suffice for tomography, the upper bound on efficient computation is the same as the best known upper bound on efficient quantum computation. Additionally, this bound holds regardless of whether the principle of \emph{causality}, which roughly states that there is no signalling from the future, is satisfied. Moreover, Ref.~\cite{landscape} showed that by slightly altering the definition of a generalised theory one can construct a theory in this modified framework that satisfies both tomographic locality and causality which achieves this upper bound, and hence can simulate any quantum computation efficiently. The current paper reviews and extends some of these results and concludes with a more nuanced version of the conjecture originally made in \cite{barrett2007information} regarding optimality of quantum computation within the generalised probabilistic theory framework.

\section{Generalised theories} \label{Section: GPTs}

A fundamental goal of any physical theory is to provide a consistent explanation of experimental data. This constitutes the core idea underlying the framework of generalised probabilistic theories, where the primitive notions are operational\footnote{Note that operationalism as a philosophical viewpoint, in which one asserts that there is no reality beyond laboratory device settings and outcomes, is not being espoused here. One should merely view the approach taken here as an operational methodology aimed at gaining insight into certain structural properties of physical theories.} in nature. Indeed, as any candidate physical theory will eventually be experimentally investigated, it should have an operational description in terms of these experiments. We will work in the circuit framework for generalised probabilistic theories developed by Hardy in \cite{hardy2011reformulating} and Chiribella, D'Ariano, and Perinotti in \cite{chiribella2010probabilistic}. The circuit framework takes inspiration from the categorical approach to quantum theory, introduced by Abramsky and Coecke \cite{abramsky2004categorical, coecke2016picturing}, in that it heavily emphasises compositionality. This compositional viewpoint also has advantages in standard quantum information, see for example Ref.~\cite{selby2020compositional}. In this manner the circuit framework  is different to the convex sets framework, an alternatively approach to generalised probabilistic theories, which was presented in \cite{barrett2007information, hardy2005probability}---although both are similar in spirit. 

Informally, a theory in this framework specifies a set of laboratory devices that can be connected together in different ways and assigns probabilities to different experimental outcomes \cite{chiribella2010probabilistic, hardy2011reformulating, barrett2007information, lee2017no}. Devices are equipped with a number of input and output ports, together with a classical pointer that can take various distinct positions. One constructs experiments by connecting the output ports of some number of devices to the input ports of others until there are no unconnected ports remaining, ensuring along the way that no cycles have been formed. When these devices are used in an experiment, the classical pointer comes to rest in a specific position, denoting the experimental outcome. Intuitively, one can think of a \emph{physical system} as passing between a devices ports. These systems come in different \emph{types}, denoted $A,B,C,\dots$. When composing devices in sequence and parallel to form experiments, the types of connected ports must match. Moreover, the set of devices is closed under sequential and parallel composition.

In the current framework, experiments---closed circuits of devices with matching types and no cycles---correspond to probabilities over the possible classical pointer positions of each constituent device in the experiment. Moreover, if a closed circuit consists of multiple disconnected circuits, then the probability assignment factorises over this decomposition. Devices with the same input and output types yielding the same probabilities in all possible experiments are identified. The equivalence class of devices with no input ports are called \emph{states}, no output ports \emph{effects}, and both input and output ports \emph{transformations}. For input type $A$ and output type $B$, these sets will be denoted $\textbf{St(B)}$, $\textbf{Eff(A)}$, and $\textbf{Transf(A,B)}$.

The correspondence between closed circuits and probabilities induces a linear structure that will be essential in the remainder of the paper. Note that states, effects, and transformations can be thought of as functions from the set of effects, states, and circuit fragments of matching type\footnote{Circuit fragments of matching type correspond to those whose unconnected ports are of the same type as the ports of the corresponding transformation.} to the interval $[0,1]$ \cite{chiribella2010probabilistic, lee2015computation}. As one can take linear combinations of functions, the set of states, effects, and transformations of a given type each respectively can be seen to generate a real vector space, in which the set of states, effects, and transformations are embedded. Note that this embedding in general results in a sub\emph{set} rather than a sub\emph{space}. Moreover, the sets of effects and transformations act linearly on the vector space generated by the set of states of the appropriate type. The vector space generated by $\textbf{St(B)}$, $\textbf{Eff(A)}$, and $\textbf{Transf(A,B)}$ will be denoted $\bold{V_B}$, $\bold{V^A}$, and $\bold{V_B^A}$ respectively.

In this work the standard assumption that the above vector spaces are all finite dimensional will be made. This corresponds\footnote{See Ref.~\cite{lee2016thesis} for a nuanced discussion of this point.} to the operational statement that there exists a finite set of \emph{fiducial} measurements whose statistics are sufficient for the tomography of arbitrary states. This implies that the vector space of state and effects are dual to one another \cite{chiribella2010probabilistic, lee2015computation}. In quantum theory, such a collection of measurements is called \emph{informationally complete} \cite{prugovevcki1977information, busch1989determination}. The correspondence between states and real vectors in the space they generate can be chosen in manifold different---yet statistically equivalent---ways \cite{masanes2013existence}. Indeed, one can always choose to represent states as real vectors whose entries correspond to the outcomes of fiducial measurements. Given a finite set of fiducial measurements, they can be combined into a single measurement consisting of a coin flip to select a measurement from the set, followed by an application of said measurement \cite{flammia2005minimal}. Hence one can always choose to represent a given state in such a theory as a real vector whose entries correspond to the probabilities of specific outcomes of such a single fiducial measurement.

A given measurement corresponds to a set of effects $\{e^r\}$ labelled by the position of the classical pointer $r$ attached to the measurement device. The probability of preparing some state $s$, observing that transformation $T_k$ has been applied, and observing outcome $r$ is (suppressing system types for readability) given by: $$e^r(T_ks) = P(r, k, s).$$
Hence the probabilities for a given circuit are calculated by performing matrix multiplication of the vectors, matrices, an dual vectors associated with the states, transformations and measurement outcomes in the given closed circuit/experiment \cite{chiribella2010probabilistic, lee2015computation}.

As mentioned previously, finite-dimensional quantum theory is an example of a generalised probabilistic theory. A quantum system is associated with a complex Hilbert space, with the system type given by the dimension of the Hilbert space. States and effects are associated with positive operators, transformations with trace non-increasing completely positive maps. A device with no input ports corresponds to what is sometimes called a `random source of quantum states', and is associated with positive operators $\{\rho_r\}$, with $r$ denoting the possible positions of the classical pointer, such that $\sum_r\mathrm{Tr}(\rho_r)=1$. When device is used in an experiment, the probability that the classical pointer takes position $r$ is given by $\mathrm{Tr}(\rho_r)$, and the quantum state prepared, conditioned on the pointer reading being $r$, is the normalised operator $\rho_r / \mathrm{Tr}(\rho_r)$. A device with no output ports is associated with a positive operator-valued measurement, that is a set of positive operators $\{E_i\}$ satisfying $\sum_i E_i=\mathbb{I}$. A device with both input and output ports is associated with a \emph{quantum instrument}, that is a set of trace non-increasing completely positive maps, one for each value of the pointer reading $r$, that sum to a trace-preserving map. Given such associations, the usual rules of quantum theory allow for calculation of probabilities for any circuit outcome. 
 
The central physical principles needed in formulation of the main results of this paper will now be formally stated. 
\begin{define}[Causality \cite{chiribella2010probabilistic}]
A theory is \emph{causal} if there exists a unique deterministic effect (the device corresponding to such an effect only has a single pointer position) for all system types. 
\end{define}
Mathematically, the principle of causality is equivalent to the statement: ``Probabilities of present experiments are independent of future measurement choices''. Hence it captures the intuitive notion that there should be ``no-signalling from the future'' \cite{chiribella2010probabilistic}. Additionally, the causality principle implies the standard ``no-signalling in space'' principle \cite{chiribella2010probabilistic, coecke2014terminality}. Moreover, the unique deterministic effect allows one to define a notion of \emph{marginalisation} for multi-partite states. Causality need not be satisfied by all generalised probabilistic theories, indeed Ref.~\cite{d'ariano2014determinism} has explicitly constructed a theory which does not satisfy the principle of causality. Quantum theory is causal, the unique deterministic effect being given by the identity matrix $\mathbb{I}_d$ for a $d$-level system.
\begin{define}[Tomographic locality for states \cite{hardy2005probability, chiribella2010probabilistic, barrett2007information}]
A theory satisfies \emph{tomographic locality for states} if the state of every composite system is uniquely specified by the statistics of measurements performed locally on each component system.
\end{define} 
Given two distinct composite states of the same type, tomographic locality implies the existence of local effects which, when applied to each individual system, give different probabilities on each state. One can generalise this principle to involve transformations as follows.
\begin{define}[Tomographic locality for transformations \cite{lee2015computation}]
A theory satisfies \emph{tomographic locality for transformations} if every transformation is uniquely characterised by local process tomography, that is, by inputting local states and performing local measurements on each component system.
\end{define}
Violation of tomographic locality (either for states or transformations) corresponds to the existence of ``global degrees of freedom'' that are not accessible at the level of local measurements. A consequence of tomographic locality\footnote{See Ref.~\cite{lee2016thesis} for a slight subtlety regarding the definition of tomographic locality in the circuit approach to generalised probabilistic theories as opposed to the convex sets approach.} is that for a transformation with input type $AB$ and output type $CD$, the corresponding real vector space has the following decomposition \cite{chiribella2010probabilistic, barrett2007information, lee2015computation}:
$$
\bold{V^{AB}_{CD}} \cong \bold{V^A} \otimes \bold{V^B} \otimes \bold{V_C} \otimes \bold{V_D},
$$
where $\otimes$ here denotes the ordinary vector space tensor product. In particular, tomographic locality for states implies that for a bipartite state of type $AC$, the corresponding vector space decomposes as $\bold{V_{AC}} \cong \bold{V_A} \otimes \bold{V_C}$. Furthermore, a transformation $T_s\in\bold{Transf(A,B)}$ is completely specified by its action on $\bold{St(A)}$ \cite{chiribella2010probabilistic}. Quantum theory satisfies both tomographic locality for states and transformations \cite{chiribella2010probabilistic}. For the case of states, one can indeed check that the real vector spaces of Hermitian operators satisfies $\bold{V_{AB}} \cong \bold{V_A}\otimes \bold{V_B}$. 
\begin{define}[$n$-local tomography for states \cite{hardy2012limited}]
A theory is \emph{$n$-locally tomographic} if the state of every composite system is uniquely specified by the statistics of measurements performed on each single system, each pair of systems, each triple of systems, $\dots$, each $n$-tuple of systems belonging to the composite.
\end{define}
As for tomographic locality, one can define an analogous principle of $n$-local tomography for transformations. Given two distinct composite states of the same type, $n$ local tomography implies the existence of $n$-local effects, consisting of conical combinations of effects that act on at most $n$-systems, which give different probabilities on each state. As $n$-local theories violate tomographic locality for $n>1$, there is a global degree of freedom which cannot be accessed by local measurements. However, as we'll see in Section~\ref{computation}, such global parameters do not grow ``too fast'' with increasing system size. 

Quantum theory defined over real---rather than complex---Hilbert spaces supplies an example of a theory that is $n=2$-local, or \emph{bi-local}, and hence does not satisfy tomographic locality. In real quantum theory, states and effects correspond to real symmetric matrices (which satisfy the same set of constraints as the standard complex quantum case discussed above), and transformations to completely positive maps that preserve the set of real symmetric matrices. We will now provide an example of two distinct transformations between two $2$-dimensional systems (sometimes referred to as \emph{rebit's} \cite{Wootters-real}), which cannot be locally distinguished in the theory and so provides a violation of tomographic locality. Consider
$$ T_1(\rho)= \frac{1}{2}\rho + \frac{1}{2}Y\rho Y, \quad \mathrm{and} \quad T_2(\rho)= \frac{1}{2}\mathbb{I} \cdot\mathrm{Tr}(\rho),$$
where $Y=\begin{pmatrix} 0& -i \\ i &0 \end{pmatrix}$ is the Pauli $Y$ matrix and $T_2$ is a measure and prepare transformation that traces out the input state and prepares the maximally mixed state, $\frac{1}{2}\mathbb{I}$. Clearly the transformation $Y\cdot Y$ is allowed in the theory as :
$$Y\rho Y= \begin{pmatrix}
0& -i \\ i &0 
\end{pmatrix}
 \begin{pmatrix} 
a& b \\ b & 1-a 
\end{pmatrix}
 \begin{pmatrix}
0& -i \\ i &0 
\end{pmatrix}
=
 \begin{pmatrix}
1-a& -b \\ -b &a
\end{pmatrix},
\quad\forall a,b \in \mathbb{R}.$$
It is easy to see that $T_1(\rho)=T_2(\rho)$ for all real symmetric matrices $\rho$ with trace one. Hence, one cannot distinguish these two transformations by inputting a single rebit and performing a single measurement on the output state. To distinguish $T_1$ from $T_2$, one has to evaluate them on one half of the Bell state $\ket{\phi^+}\bra{\phi^+}$ and perform a joint measurement of the two output systems. One has 
$$T_1\otimes{I}\left(\ket{\phi^+}\bra{\phi^+}\right)=\frac{1}{2}\ket{\phi^+}\bra{\phi^+}+\frac{1}{2}\ket{\psi^-}\bra{\psi^-},\quad \& \quad T_2\otimes{I}\left(\ket{\phi^+}\bra{\phi^+}\right)=\frac{1}{4}\mathbb{I}\otimes\mathbb{I}.$$
Performing the two-outcome measurement $\{\ket{\phi^+}\bra{\phi^+}+\ket{\psi^-}\bra{\psi^-}, \ket{\phi^-}\bra{\phi^-}+\ket{\psi^+}\bra{\psi^+}\}$, where $\ket{\phi^\pm}=\left(\ket{00}\pm\ket{11}\right)/\sqrt{2}$ and $\ket{\psi^\pm}=\left(\ket{01}\pm\ket{10}\right)/\sqrt{2}$, distinguishes these two states. Note that as 
\begin{equation} \label{global}
\left(T_1\otimes{I}-T_2\otimes{I}\right)\left(\ket{\phi^+}\bra{\phi^+}\right)=\frac{1}{4}\begin{pmatrix}
0&-1\\
1&0
\end{pmatrix}
\otimes
\begin{pmatrix}
0&-1\\
1&0
\end{pmatrix}
\end{equation}
we have $\mathrm{Tr}\left(\left(T_1\otimes{I}-T_2\otimes{I}\right)\left(E\otimes F\right)\right)=0$ for all real symmetric matrices $E,F$. Hence one cannot distinguish these two states with local measurements. Moreover, one can think of Eq.~(\ref{global}) as a global degree of freedom not acessable to local observers. It was shown by Hardy and Wooters in \cite{hardy2012limited} that one only ever needs to perform joint measurements between at most two subsystems to distinguish any two states in real quantum theory. 

\section{Computation} \label{computation} 
The class of problems a quantum computer can efficiently solve, and whose answer can be stated as either ``accept'' or ``reject'', is denoted \textbf{BQP}. Since the very beginning of Quantum Computation, much research has involved placing upper bounds on this class. Put another way, much research in quantum computing have been concerned with how large this class is. At present, the best known upper bound is that $\textbf{BQP}\subseteq\textbf{AWPP}$, where \textbf{AWPP} is a classical complexity class---known to be contained in \textbf{PP}, hence \textbf{PSPACE}---with a slightly obscure definition involving ``gap'' functions for non-deterministic Turing machines. We will provide a much more intuitive definition\footnote{See Ref.'s~\cite{morimae2015power, morimae2016modified} for another definition of \textbf{AWPP} which physicists may also find intuitive.} in section~\ref{turing machine model} which first originated in \cite{landscape}. Phrased alternatively: a quantum computer cannot efficiently solve any problem outside the class \textbf{AWPP}, but it is not known whether it can solve every problem contained within \textbf{AWPP}. Indeed, it is believed quite likely that \textbf{BQP} is a strict subset of \textbf{AWPP} \cite{landscape}.

In the following subsections, we present two different---yet equivalent---models of computation in the generalised probabilistic theory framework. The first one generalises the standard classical and quantum circuit model, and the second involves a modification of the probabilistic Turing machine model. Additionally, we review and expand previous results involving an upper bound on the power of efficient computation in a certain class of generalised probabilistic theories which first appeared in \cite{lee2015computation}. Finally, we discuss how oracles are defined in generalised theories and concludes by reviewing and extending certain oracle separation results due to \cite{lee2016deriving, lee2017oracles}.
\subsection{Circuit model} \label{circuit model}
The framework introduced in section~\ref{Section: GPTs}, in which devices can be connected together in sequence and parallel to form circuits, suggests a natural model of computation which generalises the classical and quantum circuit model of computation. In order to formally define an efficient circuit model of computation in theories belonging to the framework introduced in section~\ref{Section: GPTs}, the notion of a (polynomially sized) uniform circuit family is needed, together with a condition for a circuit to accept or reject an input. We now present such a definition, providing an intuitive explanation of each point after the formal definition. A polynomially sized uniform circuit family is a set of closed circuits $\{C_x\}$ such that:
\begin{enumerate} 
\item There is a finite gate\footnote{When discussing computation, the terms `device' and `gate' will be used interchangeably} set $\mathcal{G}$, consisting of devices, such that each circuit in the family is built from elements of $\mathcal{G}$.
\item The number of gates in the circuit $C_x$ is bounded by a polynomial in $|x|$.
\item There is a Turing machine that, acting on input $x=x_1x_2\dots x_n$, outputs a classical description of $C_x$ in time bounded by a polynomial in $|x|$.
\item For each type of system, there is a fixed choice of basis, relative to which transformations are associated with matrices. Given the matrix ${M}$ representing (a particular outcome of) a gate in $\mathcal{G}$, a Turing machine can output a matrix $\widetilde{{M}}$ with rational entries, such that $ | ({M} - \widetilde{{M}})_{ij} | \leq \epsilon$, in time polynomial in $\log(1/\epsilon)$. 
\end{enumerate}
Regarding item $1$, for a specific theory in our framework it may not be the case that gates acting only on bipartite and single systems are universal for computation, as is the case for quantum computation. Thus for any $m,k$, a circuit might involve gates with $m$ input systems and $k$ output systems. In general, it may be that no finite gate set is universal for computation. Nonetheless, we demand as a requirement of uniformity that any uniform circuit family is built from elements of a finite gate set\footnote{One may alternatively consider a uniformity condition in which the number of allowed gates increases with growing circuit size, as is the case in \cite[Section 3.3~A]{de2014computation}.}. A consequence is that the number of distinct system types appearing in a given uniform circuit family is also finite.  

Regarding item $2$ and $3$, as we saw in section~\ref{Section: GPTs}, not every generalised probabilistic theory satisfies the principle of causality, in which case a circuit does not have a `preferred' direction. Hence there is no reason to assume---as is standardly done for classical and quantum circuits---that a circuit must have the form of a number of gates acting on some input, where the input preparation encodes the problem instance. Instead, we permit the entire circuit to encode the problem instance, defining a circuit family as a set $\{C_x\}$ with the stipulation that each circuit is indexed by a classical string $x=x_1x_2\dots x_n$. A circuit family is polynomial-size if the number of gates is bounded from above\footnote{By padding with identity transformations, one can always ensure that the number of gates is exactly specified by a polynomial in $|x|$.} by a polynomial in $|x|$. Note that there are formulations of quantum circuits in which the entire circuit encodes the problem instance rather than just the input states, see section $2.2$ of Ref.~\cite{bremner2010classical} for a specific example.

Regarding item $4$, the final requirement for a circuit family to be uniform manifests as a constraint on the entries of the matrices representing the transformations that appear in the finite gate set. If such a constraint is not made, it may be possible to smuggle hard to compute quantities into the computation through the matrix entries. Note that such a constraint is required even in the classical case. Indeed, consider a coin whose probability of landing heads up is equal to Chaitin's constant \cite{chaitin1975theory}, which, informally, represents the probability that a given Turing machine will halt on a randomly chosen input. By flipping such a coin multiple times, once could extract the first $N$ bits of this probability. Using such an approximation one could solve the halting problem for all inputs of size up to $N$ in some fixed amount of time. This is believed not to be possible, hence there should exist some constraint on which probabilities are accessible computationally. We can motivate such a constraint by recalling that gates correspond to operational devices; an experimenter with access to devices governed by some generalised probabilistic theory may only be able to characterise them tomographically to finite precision---and the features of a probabilistic theory should not be sensitive to precision issues which are inaccessible to experiment. We formalise this constraint as follows: there must exist some fixed choice of basis for $\bold{V_A}$ for each system $A$, such that a Turing machine can efficiently compute approximations to the entries of the matrices relative to these bases. We require that for any matrix entry $({M})_{ij}$, and any $\epsilon$, a Turing machine can output a rational number, within $\epsilon$ of $({M})_{ij}$, in time bounded by a polynomial in $\log(\frac{1}{\epsilon})$.  

Denoting the string of observed outcomes by $z$, the final output of the computation will be given by a function $a(z)\in \{0,1\}$. We say that a run of the experiment accepts an input string $x$ if the outcome string $z$ of the circuit $C_x$ satisfies $a(z)=0$. We require the existence of a Turing machine that computes $a$ in time polynomial in the length of the input $|x|$. This final requirement is needed to ensure that the calculation of whether a given circuit outcome accepts or rejects should not itself be able to solve a computationally hard problem. The probability that a computation accepts the input string $x$ is thus
$$P_x({\mathrm{accept}}) \,=\! \sum_{z|a(z)=0}\!\!P(z),$$ where the sum ranges over all possible outcome strings $z$ of the circuit $C_x$ for which $a(z) = 0$.  

In the quantum case, one usually says a given computation accepts its input if a measurement of the first outcome qubit in the computational basis yields $\ket{0}$. This is due to the fact that the application of any non-deterministic quantum operation at an intermediate stage in the computation can always without loss of generality be replaced by a unitary operation (potentially acting on some extra auxiliary qubits) followed by a projective measurement at the end of the computation. Hence, the only outcomes, i.e. non-trivial classical pointer positions, generated by the computation appear on measurements performed at the end of the circuit. In general it is not the case that one can postpone non-deterministic operations (that is, operations represented by a device with more than a single classical pointer position) to the end of a circuit in an arbitrary generalised probabilistic theory.\footnote{\color{black} If the theory under consideration satisfies the \emph{Purification Principle} of \cite{chiribella2010probabilistic}, then non-deterministic measurements can indeed be postponed to the end of a computation as in such theories any non-deterministic operation can be ``dilated'' to a reversible operation followed by a measurement. In fact, the Purification Principle is equivalent to the existence of such a reversible dilation for every operation in the theory \cite{chiribella2010probabilistic}.\color{black}} Hence the need for the generalised acceptance condition outlined above. 

Given the notion of a uniform circuit family and our generalised acceptance condition, we can now formally define the class of problems that can be efficiently solved in an arbitrary generalised probabilistic theory.
\begin{define}[Class of efficient computation in a generalised probabilistic theory \cite{lee2015computation}] \label{BGP}
For a theory $\bold{G}$, a language $\mathcal{L}$ is in the class $\bold{BGP}$ if there exists a poly-sized uniform family of circuits in $\bold{G}$, and an efficient acceptor, such that
\begin{enumerate}
\item $x\in\mathcal{L}$ is accepted with probability at least $\frac{2}{3}$. 
\item $x\notin\mathcal{L}$ is accepted with probability at most $\frac{1}{3}$. \end{enumerate}  
\end{define}
The constants in the above definition can be chosen arbitrarily as long as they are bounded away from a half by an inverse polynomial in $x$ \cite{lee2015computation}. 

The following theorem was proved in \cite{lee2015computation}:
\begin{thm}[Theorem 3.4.1 in \cite{lee2015computation}] \label{bgp in awpp}
For any generalised probabilistic theory $\bold{G}$ satisfying tomographic locality, the following holds: $$\bold{BGP}\subseteq\bold{AWPP}\subseteq\bold{PP}\subseteq\bold{PSPACE}.$$
\end{thm} 
Note that the principle of causality was not required to derive the above bound. Once the appropriate definitions are in place, the proof of theorem~\ref{bgp in awpp} is a fairly straightforward extension of similar proofs in the quantum case \cite{fortnow1998complexity}. We will now give an intuitive overview of how the operational assumptions underlying the generalised probabilistic theory framework together with tomographic locality allow for the derivation of the bound from theorem~\ref{bgp in awpp}.

First, the proof relies on the fact that transformations in a generalised probabilistic theory are linear, and hence have a matrix representation. As we saw in section~\ref{Section: GPTs} this linear structure arose from the intuitive requirement that a physical theory should be able to give probabilistic predictions about the occurrence of possible outcomes. Second, a further requirement of the proof is the ability to compute efficiently the entries in the matrices representing the transformations applied in parallel in a specific circuit. Recall from section~\ref{Section: GPTs}, that in a tomographically local theory the matrix corresponding to transformations applied in parallel can be easily calculated by taking the tensor product of the matrices representing each individual transformation. As the matrices for individual gates satisfy the uniformity condition, the elements of tensors products of these matrices do as well. By foliating the circuit so that only a single non-trivial gate acts in any given foliation, one sees that the outcome probability
of the circuit corresponds to the multiplication of matrices consisting of tensor products of these gates with identity transformations on systems on which they do not act \cite{lee2015computation}. 
Generically, this is not the case without tomographic locality---as the tensor product structure may not hold.  

If a transformation from $\bold{A}$ to $\bold{B}$ acts only on a subsystem of $\bold{AC}$, there might be no simple relation between the linear map $\bold{{St}(AC)}\rightarrow\bold{{St}(BC)}$ and the arising from the action of the transformation when it is applied to system $\bold{A}$ on its own, or indeed to a joint system $\bold{AD}$, for some other system $\bold{D}$. There may thus be no efficient way of computing matrix elements corresponding to a transformation considered as part of a circuit of arbitrary size. Recall that one can think of a violation of tomographic locality as akin to the existence of ``global degrees of freedom'' not accessible via local measurements. As the uniformity condition only imposes constraints on the matrices associated to gates from the finite gate set $\mathcal{G}$, if there doesn't exist an efficient procedure for computing the matrices associated to the parallel compositions of gates from $\mathcal{G}$ one could in principle encode answers to computationally hard problems in these global degrees of freedom. While tomographic locality implies such an efficient procedure, it may not be the simplest principle to do so. 

Real Hilbert space quantum theory, discussed at the end of section~\ref{Section: GPTs}, provides an example of a theory without tomographic locality for which the bound of theorem~\ref{bgp in awpp} still holds, since there is an efficient way of calculating the relevant matrix entries. One can in fact show that any theory which satisfies $n$-local tomography, for $n$ a fixed constant, also satisfies the bound of theorem~\ref{bgp in awpp}.

In our original set up, our uniformity assumption ensured that the entries of each matrix from our finite gate set could be efficiently approximated by a Turing machine, i.e. a `classical' computer. One could imagine the classical computer being fed the probabilities of fiducial measurement outcomes, from which it efficiently calculates the matrix corresponding to this transformation in some fixed basis. This statement just sums up the fact that once one has collected all the necessary statistics, one can ``easily'' output the state (or transformation) they describe. Without such an commitment, one could not do tomography. Hence to output an approximation of the matrix entry of some transformation, a classical computer has to hold in its memory the probabilities of all fiducial measurement outcomes. From this it can easily approximate (to some degree of error) the required matrix entry. 

Now, if we assume tomographic locality then the classical computer only needs to store a finite number of these fiducial probabilities. These are just the fiducial outcomes for the system on which the non-identity transformation acts, as the matrix entries for the tensor product of this transformation with identities are either zero, or just the entries of the non-identity transformation itself (from uniformity of the circuit, the computer can efficiently compute on which system the non-identity transformation acts). In the case of $n$-local tomography, for $n$ a fixed constant, the number of measurements needed to specify a given transformation grows with system size, but it doesn't grow ``too fast.'' For a transformation acting on $N$ systems, the number of measurements required is
\begin{equation} \label{increasing measurement statistics}
k\left(\frac{N!}{n!(N-n)!}\right),
\end{equation}
where $k$ is the largest integer such that the outcome of $k$ fiducial measurements is sufficient to determine any state of a given system. Note that as only finitely many systems appear in a given circuit family (item $1$ from the statement of uniformity), $k$ is a constant. To leading order, Eq.~(\ref{increasing measurement statistics}) goes as $cN^n$, with $c$ and $n$ constant. \color{black} Hence the classical computer needs to store a polynomially growing set of fiducial measurement outcomes; it needs a poly-size memory. However, this machine only requires at most poly-time to run, acting in some sense as a transducer which writes the description of the matrix on an output tape. Hence a standard poly-size, poly-time Turing machine suffices. \color{black}

Given the above argument, the remainder of the proof of theorem~\ref{bgp in awpp} in \cite{lee2015computation} goes through verbatim. We have thus provided a sketch proof of the following theorem.
\begin{new}[Extends Theorem 3.4.1 in \cite{lee2015computation}]
For any generalised probabilistic theory $\bold{G}$ satisfying $n$-local tomography, for $n$ a fixed constant, the following bound holds holds: $$\bold{BGP}\subseteq\bold{AWPP}\subseteq\bold{PP}\subseteq\bold{PSPACE}.$$
\end{new} 
If a theory does not even satisfy $n$-local tomography for any fixed $n$, then the number of fiducial outcomes needed to specify a transformation could grow exponentially in system size. A poly-size Turing machine would not be able to store all of these in its memory, hence it appears the bound will not hold in these situations.

\subsection{Oracles} \label{oracles}
In quantum computing, \emph{oracles} are a vital component of many of the known computational speed-ups over classical computing \cite{bennett1997strengths, grover1997quantum, nielsen2010quantum}. In general, an oracle provides access to some function $f:\{1,\dots, N\}\rightarrow\{0,1\}$ and one is usually concerned with the number of queries to this oracle needed to solve a certain problem. In quantum computing an oracle corresponds to a controlled unitary transformation $U_f$ (in fact, to a family of controlled unitaries---one for each problem size) which acts on the computational basis $\{\ket{x}\}$ as
$$
U_f \ket{x}\ket{y}= \ket{x}\ket{y\oplus f(x)}.
$$
By performing measurements on the target system, represented above by the $\ket{y}$ state, one can learn about the value of $f$ on specific inputs.

However, a generic generalised probabilistic theory does not have sufficient structure to define such an oracle \cite{lee2015computation}. Without one, it is difficult to compare the computational abilities of different theories and hence to assess how computational power depends on different physical principles. In Ref.~\cite{lee2015computation} a rudimentary oracle model was defined for all theories satisfying the principle of causality. It was shown that, relative to this oracle model, the class \textbf{NP} was not contained in any theory satisfying tomographic locality and causality. A major drawback of this model however, was the fact that it cannot be queried in superposition---that is, in a non-classical manner. However, it was shown in Ref.~\cite{lee2016generalised} that in theories satisfying certain natural physical principles\footnote{Namely: causality, purification, strong symmetry, and the requirement that the parallel composition of pure states is also a pure state. See \cite{lee2016generalised} for more information.} reversible controlled transformations exist and provide a natural generalisation of quantum oracles. 

Using this oracle model, Ref.~\cite{lee2017oracles} proved that the best achievable lower bound on the number of quantum queries needed to solve certain query problems is not optimal in the space of all generalised probabilistic theories. An example of such a query problem is \texttt{PARITY}, a generalisation of Deutsch's problem \cite{nielsen2010quantum} which asks for the parity of a function $f:\{1,\dots,N\}\rightarrow\{0,1\}$---that is, the value $f(1)\oplus\cdots\oplus f(N) \text{ mod }2$---where $N$ is taken to be a constant. On a classical computer, $N$ queries are needed to solve this problem, but $\lceil{N/2}\rceil$ queries to a quantum oracle suffice to determine the parity. In fact, $\lceil{N/2}\rceil$ queries to a quantum oracle is optimal; a quantum computer cannot determine the parity using fewer queries \cite{meyer2011uselessness}. However, if a theory exhibits certain type of post-quantum, or higher-order \cite{lee2016higher, sorkin1994quantum, barnum2017ruling, barnum2015entropy, niestegge2016quantum}, interference---which we discuss in more detail below---then a mathematically achievable lower bound to the number of queries needed to determine the parity is $1$ \cite{lee2017oracles}---although the physical principles required to reach this bound is unknown in general. 

Related work in Ref.~\cite{lee2016deriving} derived Grover's lower bound to the search problem from simple physical principles\footnote{The same principles required to define oracles in generalised theories mentioned in footnote 11.}. The search problem asks one to find a certain ``marked item'' from among a collection of items in an unordered database. The only access to the database is through an oracle; when asked if item $i$ is the marked one, the oracle outputs ``yes'' or ``no''. The figure of merit in this problem is how the minimum number of queries required to find the marked item scales with the size of the database. It was shown that, asymptotically, post-quantum, or higher-order, interference does not provide an advantage over quantum theory in this case, as the number of queries needed to find the marked item in both cases scales as the square root of the database size. For increasing database size $N$, the lower bound to the number of queries scales as $\sqrt{N}$ in both cases. Hence, post-quantum interference is not, in general, a resource for post-quantum computer. Although post-quantum interference may provide a constant speed-up for certain problems, this advantage does not lead to improved scaling with problem size.

The above results were originally derived from the same collection of physical principles (which are outlined in Footnote 11). However, those principles were only used to derive a specific, and physically intuitive, representation for states in those theories from first principles. Once the representation is specified, the above results can be proven directly, without reference to the physical principles. Moreover, while those principles are sufficient to derive such a representation, they do not appear to be necessary \cite{dakic2014density}. Moreover, there is general interest in theories with such representations \cite{dakic2014density, gogioso2018density, hefford2020hyper}. Hence it is interesting to understand computational consequences of the representation alone, rather than computational consequences of sufficient set of physical principles that imply it.

To describe the above mentioned representation, we first need to define higher-order interference. The definition of higher-order interference that we present here takes its motivation from the set-up of multi-slit interference experiments. In such experiments a particle (a photon or electron, say) passes through slits in a physical barrier. By blocking some of the slits and repeating the experiment many times, one can build up an interference pattern on a screen placed behind the physical barrier. Informally, a theory has ``$n$th order interference'' if one can generate interference patterns in an $n$-slit experiment which cannot be created in any experiment with only $m$-slits, for all $m<n$.

It was first shown by Sorkin \cite{sorkin1994quantum,sorkin1995quantum} that---at least for ideal experiments \cite{sinha2015superposition}---quantum theory is limited to the $n=2$ case. That is, the interference pattern created in a three---or more---slit experiment \emph{can} be written in terms of the two and one slit interference patterns obtained by blocking some of the slits. 

Given $N$ slits, labelled $1, \dots, N$, these transformations will be denoted $P_I$, where $I \subseteq \{1, \dots, N\}$ corresponds to the subset of slits which are not closed. In general one expects that $P_I P_J = P_{I\cap J}$, as only those slits belonging to both $I$ and $J$ will not be closed by either $P_I$ or $P_J$. Thus we think of these transformations as corresponding to projectors (i.e. idempotent transformations $P_IP_I=P_I$). Instead of working directly with these physical projectors, it is mathematically convenient to work with the (generally) unphysical transformations corresponding to projecting onto the ``coherences'' of a state. Consider the example of a qutrit in quantum theory, the projector $P_{\{0,1\}}$ projects onto a two dimensional subspace:
\[ P_{\{0,1\}}::\left(\begin{array}{ccc} \rho_{00} & \rho_{01} & \rho_{02} \\ \rho_{10} & \rho_{11} & \rho_{12} \\\rho_{20} & \rho_{21} & \rho_{22}   \end{array}\right)\mapsto \left(\begin{array}{ccc} \rho_{00} & \rho_{01} & 0 \\ \rho_{10} & \rho_{11} & 0 \\0 & 0 & 0   \end{array}\right)\]
whilst the coherence-projector $\omega_{\{0,1\}}$ projects only onto the coherences in that two dimensional subspace:
\[ \omega_{\{0,1\}}::\left(\begin{array}{ccc} \rho_{00} & \rho_{01} & \rho_{02} \\ \rho_{10} & \rho_{11} & \rho_{12} \\\rho_{20} & \rho_{21} & \rho_{22}   \end{array}\right)\mapsto \left(\begin{array}{ccc} 0 & \rho_{01} & 0 \\ \rho_{10} & 0 & 0 \\0 & 0 & 0   \end{array}\right).\]
That is, $\omega_{\{0,1\}}$ corresponds to the linear combination of projectors: $P_{\{0,1\}}-P_{\{0\}}-P_{\{1\}}$.

There is a coherence-projector $\omega_I$ for each subset of slits $I \subseteq \{1,\dots, N\}$, defined in terms of the physical projectors:
$$\omega_I:=\sum_{\tilde{I}\subseteq I}(-1)^{|I|+|\tilde{I}|}P_{\tilde{I}}.$$ 

If we demand that any state (indeed, any vector in the vector space generated by the states) in a theory can be decomposed in a form reminiscent of a rank $k$ tensor:
\begin{equation} \label{coherence decomposition}
|s)=\sum_{I,|I|=1}^k \omega_I|s)=\sum_{I,|I|=1}^k |s_I),
\end{equation}
then that theory has maximal order of interference $k$. This decomposition can be thought of as a generalised superposition, as it manifestly describes the coherences between different subsets of perfectly distinguishable states (the analogue of a basis in quantum theory) present in a given state. 

Given this representation, we can now state the main results of \cite{lee2017oracles} and \cite{lee2016deriving} without mention of physical principles.

\begin{new}[Extends Theorem 3.0.4 in \cite{lee2017oracles} and Theorem 1 in \cite{lee2016deriving}]
In any theory were states can be represented as a rank-k tensor as in Eq.~\ref{coherence decomposition}, for $k$ a constant, the \texttt{PARITY} problem for a function $f:\{1,\dots,N\}\rightarrow\{0,1\}$ requires a minimum of $\lceil{N/k}\rceil$ queries to solve, and the search problem on a database of size $N$ can be solved with $\Omega(\sqrt{N/ k})$ queries.
\end{new}

Hence, theories where states can be represented as generalised superpositions do not provide a computatonal advantage over quantum theory. Although generalised superpositions may provide a constant speed-up for certain problems, this advantage does not lead to improved scaling with problem size. 

\subsection{Turing machine model} \label{turing machine model}
The original formulation of quantum and classical computation was in terms of (quantum and deterministic) Turing machines, rather than circuits. Can a model of computation in an arbitrary theory be formulated that utilises Turing machines? In this section, building on work in \cite{landscape}, we present such a generalised Turing machine model.     

We start by introducing a new type of Turing Machine, generalising the standard notion of a probabilistic Turing machine. In this new model, transitions can occur with quasi-probabilistic weights---rather than the standard transition probabilities occuring in probabilistic Turning machines---with the constraint that the total weight of transitions from a given state must sum to $+1$. We refer to this model as an \emph{Affine Turing Machine} \cite{landscape}. It was shown in \cite{landscape} that the class of problems which can be efficiently solved by this model with bounded error is exactly equal to the class \textbf{AWPP}. Hence, by replacing probabilities with quasi-probabilities, one gets from the well known class \textbf{BPP} to \textbf{AWPP}, thus providing an intuitive interpretation of \textbf{AWPP}.

More formally, an \emph{Affine Turing Machine} (AffTM) is defined to be a non-deterministic Turing machine in which every transition is associated with a real-valued---not necessarily positive---\emph{weight}. The weight of a given computational branch corresponds to the product of the weights of the transitions involved.	 It's required that for each symbol being read by the tape head, the total weight of transitions from a given (non-halting) state must be $+1$.

Given the above, the class of languages an AffTM can efficiently decide with bounded error can now rigorously be defined.
First, given an AffTM $\mathbf M$ whose branches all halt in a finite number of steps, define the \emph{acceptance weight} $\alpha_{\mathbf M}(x)$ of $\mathbf M$ on an input $x$ to be the total weight of the accepting paths on input $x$.
An AffTM $\mathbf M$ is said to be \emph{proper} if $0 \le \alpha_{\mathbf M}(x) \le 1$ for all inputs. It decides a language $L$ with \emph{bounded error} if $\tfrac{2}{3} \le \alpha_{\mathbf M}(x) \le 1$ for $x \in L$, and $0 \le \alpha_{\mathbf M}(x) \le \tfrac{1}{3}$ for $x \notin L$. An AffTM is \emph{efficient} if the number of transitions in any computational path on an arbitrary input $x$ is bounded from above by some polynomial in $|x|$. Given these definitions, we can now formally state the relation between AffTMs and \textbf{AWPP}.
\begin{thm}[Affine Turing machine characterisation of \textbf{AWPP} \cite{landscape}]\label{afftm equal to awpp}
The class of languages decided with bounded error by some efficient proper AffTM is equal to $\bold{AWPP}$.
\end{thm} 
This correspondence
will be used to provide a Turning machine model of computation for an arbitrary theory---equivalent to the uniform circuit family model discussed in section~\ref{circuit model}---in terms of an AffTM satisfying one further constraint.

Fix a theory and consider a language $L$ that can be efficiently decided by it with bounded error. We saw in section~\ref{circuit model} that $L\in\textbf{AWPP}$, hence there exists a proper AffTM that can efficiently decide $L$ with bounded error. Write the ``affine vector'' of this AffTM at a given time as a quasi-probability distribution over all configurations---with the term in quasi-distribution associated to a given configuration corresponding to the weight of that configuration---of the AffTM at that point in the computation.  As the uniform circuit family in the theory simulates this AffTM, the real vector corresponding to the state of the theory at a given point in the circuit is an alternate representation of the affine vector at the equivalent point in the AffTM computation.

Recall from section~\ref{Section: GPTs} that in each theory the set of fiducial measurements can be combined into a single informationally complete measurement in such a way that the each state corresponds to the vector listing the outcome probability for each effect in this informationally complete measurement. Hence, the Euclidean norm, or $2$-norm, of each state must be bounded\footnote{This follows from the fact that the state with outcome probability $1$ for a given effect (and hence probability $0$ for every other effect due to normalisation of probabilities) \emph{magorizes} \cite{marshall1979inequalities} every other state. The $2$-norm of this state is clearly $1$. As the $2$-norm is a Schur-convex function \cite{marshall1979inequalities} the $2$-norm of every other state is hence bounded from above by $1$.} from above by $1$. Thus the $2$-norm of each affine vector---that is quasi-distribution over configurations---of the AffTM simulating a given circuit in the theory must also be bounded above by $1$.

Hence, for a given language efficiently decided with bounded error by a specific theory \textbf{G}, there exists an efficient proper AffTM for which the Euclidean norm of each of its affine vectors is upper bounded by $1$, which also decides this language with bounded error. This provides an alternate characterisation of \textbf{BGP} to the one provided in definition~\ref{BGP} which makes use of Turning machines rather than uniform circuit families.  

\section{Achieving the upper bound: free vs. non-free theories}

A natural question is whether there exits a theory such that the bound of theorem~\ref{bgp in awpp} is achieved. In fact, Ref.~\cite{landscape} has provided a complexity-theoretic argument that suggests this is unlikely. Despite this, by slightly modifying the definition of what constitutes a generalised probabilistic theory, one can indeed construct a theory within this altered framework---satisfying both tomographic locality and causality---in which the class of efficiently solvable problems exactly equals \textbf{AWPP} \cite{landscape}. As we will see below, rather than assigning probabilities to any experiment composed of laboratory devices, a theory in the modified framework only assigns probabilities to certain \emph{allowed} experiments \cite{landscape}. This result constitutes a converse to theorem~\ref{bgp in awpp} and provides an intuitive interpretation of \textbf{AWPP}, which can be thought of as the class of all problems efficiently solvable by tomographically local physical theories. The theory constructed in Ref.~\cite{landscape} has the maximum computational power consistent with tomographic locality. In a sense, one can think of it as the analogue of a PR-box---which exhibits the strongest non-local correlations consistent with the no-signalling principle---for computation.

The standard definition of a generalised theory, outlined in section~\ref{Section: GPTs}, holds that a theory specifies a set of experimental or laboratory devices which can be composed together in sequence and parallel to form closed circuits and assigns a probability distribution over the possible outcomes of each closed circuit. Additionally, the set of devices---and device outcomes---is closed under such sequential and parallel composition. Ref.~\cite{landscape} referred to such theories as \emph{free} generalised probabilistic theories.  One can consider a modified definition of a theory, which specifies a set of devices, a set of \emph{allowed} closed circuits which can be built from those devices, and assigns a probability distribution over the outcomes of allowed closed circuits. Note that probability distributions are only assigned to the set of \emph{allowed} closed circuits specified by the theory. Such theories were referred to by Ref.~\cite{landscape} as \emph{non-free} theories. In non-free theories, states, transformations, and effect are still represented by vectors in a real vector space, matrices acting on this space, and vectors in the dual space respectivley \cite{lee2016thesis}. 

Before proceeding, it is natural to ask whether non-free theories constitute a sensible class operationally defined theories. Indeed, a standard assumption made in quantum information theory is that any circuit of unitaries can be constructed. The fact that non-free theories explicitly deny this assumption may strike some as unnatural. However, non-free theories are not as unmotivated as they may seem. All that is involved in an experiment is to start to some process in motion, and at some point intervene to make an observation of some kind. Suppose that ultimately, evolution in our universe is governed by the equations of the Standard Model. Then the Hamiltonian of the Standard Model is fixed, and one does not have the option of changing it in order to realise an arbitrary Hamiltonian evolution. The fact that certain subsystems can be identified (say, some ions in some trap, say) and their evolution controlled in an arbitrary way by placing them into a suitable environment (by applying lasers to the ions, say), is rather special, and it is as much to do with the initial condition as with the laws of the universe. Thus the statement that an experimenter can start processes in motion, intervening at some point to make an observation, and that the process can be abstractly represented as a sequence of gates is not at odds with the assertion that there may be no physical process corresponding to an arbitrary sequence of the same gates. Hence, non-free theories are not unmotivated if one takes the viewpoint that a physical theory corresponds both to a consistent account of experimental data and to which experiments are implementable in principle.

Given the definition of non-free theories, Ref.~\cite{landscape} proved the following theorem.
\begin{thm}  \label{non-free equal to awpp}
There exists a non-free theory \textbf{G}, satisfying tomographic locality and causality, such that $\textbf{BGP}=\textbf{AWPP}$. 
\end{thm}
In Ref.~\cite{landscape}, it was shown that uniform poly-size circuits, in which the gates are certain affine transformations, can simulate---and be simulated by---a given AffTM. Hence, each language decided with bounded error by some efficient proper AffTM can also be decided with bounded error by a uniform circuit family built from a certain finite set of affine gates. The non-free theory of the above theorem is constructed from the collection of affine circuits which each AffTM, hence all of \textbf{AWPP}. The non-trivial part of the proof of theorem~\ref{non-free equal to awpp} is showing the non-free theory constructed in this manner satisfies tomographic locality.

\section{A new conjecture} \label{section: a new conjecture}

As \textbf{BQP} is believed to be strictly contained in \textbf{AWPP} \cite{landscape}, theorem~\ref{non-free equal to awpp} constitutes evidence against a conjecture originally made in Ref.~\cite{barrett2007information} which posited that a quantum computer may be able to simulate computation in any generalised probabilistic theory with at most polynomial overhead. The distinction between free and non-free theories appears to capture an important aspect of the computational abilities of generalised probabilistic theories. The crucial distinction between free and non-free theories is that transformations in free theories are closed under composition, implying a bound on the set of states. Indeed, this is the intuitive rationale behind the constraint that the affine vectors of any AffTMs simulating a free theory must have $2$-norm bounded by $1$; as the state space is bounded, the length of the affine vector cannot increase beyond a certain limit over the course of the computation. This need not be the case in non-free theories. Could a quantum computer exploit this fact and efficiently simulate computation in all tomographically local free theories? If such a conjecture was borne out, it could shed light on which physical and structural features give rise to the quantum computational speed-up. The conjunction of theorem~\ref{bgp in awpp} and theorem~\ref{non-free equal to awpp} suggests the following refinement of the conjecture originally made in Ref.~\cite{barrett2007information}.
\begin{conjecture}[Refinement of Conjecture 2 from \cite{barrett2007information}]
A quantum computer can simulate computation in any free theory satisfying $n$-local tomography, for $n$ a fixed constant, with at most polynomial overhead.
\end{conjecture} 
\noindent Indeed, the results outlined in section~\ref{oracles} provide weak evidence for the above conjecture. Moreover, previous results have shown that quantum theory can simulate reversible computation in Boxworld \cite{gross2010all} and in a theory where wires in computational circuits are represented by $d$ dimensional Bloch balls, for $d \neq 3$, rather than the $3$ dimensional Bloch ball of qubits in quantum computation \cite{krumm2019quantum}. Both of these simulated settings satisfy tomographic locality.

\section{Discussion and conclusion}  

This paper has reviewed and extended recent results which have explored some connections between computation and physical principles in the framework of generalised probabilistic theories \cite{lee2016generalised, lee2015computation, landscape, lee2015proofs, lee2016deriving, lee2016higher, lee2017oracles}. The main focus has been on understanding how simple physical principles bound the power of different computational paradigms, a first step towards understanding the source of quantum theories computational power \cite{lee2019heart}. In section~\ref{circuit model}, we extended a result of Ref.~\cite{lee2015computation} by showing that in any theory  satisfying $n$-local tomography the class of problems that can be solved efficiently is contained in \textbf{AWPP}---the best known bound on the power of quantum computation \cite{fortnow1998complexity}. 
 
These results raised the question of whether quantum theory is powerful for computation in the space of all theories. Given a specific theory, under what conditions can computation in this theory be simulated by a quantum computer? Put differently, can \textbf{BQP}, the class of problems a quantum computer can efficiently solve, be characterised in terms of physical principles alone? Such a characterisation would deepen our understanding of quantum computation and its ultimate limitations. One can interpret the derivation of Grover's quadratic lower bound to the search problem \cite{lee2016deriving}, discussed in section~\ref{oracles}, as the analogue of the derivation of Tsirelson's bound on quantum correlations from physical principles \cite{pawlowski2009information, fritz2013local}. A theory-independent characterisation of \textbf{BQP} would then amount to the analogue of a characterisation of the entire set of quantum correlations in terms of natural principles.   

While the results discussed in this paper may have deepened our understanding of quantum computers and their limits, they have not explicitly resulted in any practical applications in the same way that studying Boxworld type correlations led to the development of device-independent cryptography \cite{barrett2005no}. Can studying computation in general theories different from quantum theory result in practical applications? One potential avenue for this is blind and verified delegated computation \cite{blind}. 
 
Consider the situation where a computationally bounded client wants to delegate her computation to a server with access to a full quantum computer. The protocol for blind computation provided in \cite{blind} ensures that the client can have a server carry out a quantum computation for her such that the client's inputs, outputs, and computation remain perfectly private. Hence, a malicious server cannot learn any of the client's information, and all the client needs to be able to do is prepare single qubit states and send them to the server.\footnote{\color{black}It should be pointed out that this problem becomes non-trivial only when the communication between verifier and prover involves communicating quantum states, as in the protocol by Ref~\cite{blind} discussed here. When the communication between a single prover and verifier is classical, then all standard results in interactive proof systems  hold, since they apply to arbitrary provers.\color{black}} Moreover, the security of this protocol has been shown to follow from the no-signalling principle \cite{blind3}. However, while the server cannot learn the client's computation, they can still tamper with it by deviating from the client's instructions\footnote{Indeed, the server could simply refuse to perform the client's computation. Such a refusal is immediately obvious to the client. As nothing can be done to force the server to perform the computation, we only consider situations in which the server does carry out a computation. It is then up to the client to verify if the performed computation is the one specified.}. Using a scheme introduced by Ref.~\cite{blind}, the client can detect any deviations made by the server with probability exponentially close to one. 

However, the correctness of this verification protocol rests on the assumption that the server can only deviate from the specified instructions by using \emph{quantum} dynamics. The client may not be able to detect deviations which use \emph{post-quantum} dynamics. Hence, as was the case for quantum key distribution before the work of Barrett, Hardy, and Kent \cite{barrett2005no}, it is not clear if delegated quantum computation is secure against post-quantum attacks. This raises the question of whether the correctness of this verification protocol can be established directly from physical principles. Indeed, a derivation of \textbf{BQP} from physical principles would lay the groundwork for a proof of the correctness of delegated computation solely from first principles.

\color{black} One can already provide some initial analysis using known results in the literature. For instance, as the protocol of \cite{blind} requires the client to prepare and communicate single qubit states, one can wonder what additional requirements are needed to ensure the client can detect deviations when the server is not assumed to employ quantum dynamics a priori. In order for it to make good operational sense, we assume the theory used by the server must belong to the framework discussed here. Ref.~\cite{de2014computation} has shown that quantum theory is the unique generalised probabilistic theory theory where the local systems are qubits, global systems obey tomographic locality, and where there exists at least one continuous interaction between systems. Hence, if one assumes such structure, then the standard correctness of delegated computation discussed above applies. We have replaced the requirement that the server is bound by quantum theory by the assumption of tomographic locality, and the existence of a reversible, continuous interaction between subsystems---as long as local systems are qubits. In fact, one can go further. Recent work in Ref.~\cite{krumm2019quantum} has shown that if local systems are $d$-dimensional Bloch balls (a qubit, for instance, lives in a 3 dimensional Bloch ball), tomographic locality and causality are satisfied, and the global transformations form a closed, connected matrix, then the only theory with non-trivial interactions is when $d=3$. That is, when local systems are qubits. All other cases only have local transformations, and hence can be simulated on a classical computer.  

In the above two instances we have been able to replace the requirement that the server be bound by quantum theory by other---and in the second case, weaker---assumptions. If one has a full derivation of \textbf{BQP} from physical principles, then one could perhaps replace the requirement that the server be bound by quantum theory in the correctness proof by the requirement the server satisfy these physical principles. This would place the correctness and security of delegated computation on par with physical principles, in the same way that Barrett, Hardy, and Kent \cite{barrett2005no} placed security of quantum key distribution on par with the no-signalling principle.

\color{black}

\section*{Acknowledgements}
The author thanks Jonathan Barrett and John Selby for useful discussions. This project was funded by the EPSRC through the National Quantum Technology Hub in Networked Quantum Information Technologies and the UCL doctoral prize fellowship (project number: 534936).

\bibliographystyle{ieeetr}
\bibliography{library}

\begin{thebibliography}{10}

\bibitem{Peres}
A.~Peres, ``Quantum theory: Concepts and methods,'' {\em Kluwer Academic,
  Boston}, 1995.

\bibitem{hardy2005probability}
L.~Hardy, ``Probability theories with dynamic causal structure: a new framework
  for quantum gravity,'' {\em arXiv preprint gr-qc/0509120}, 2005.

\bibitem{hardy2011reformulating}
L.~Hardy, ``Reformulating and reconstructing quantum theory,'' {\em arXiv
  preprint arXiv:1104.2066}, 2011.

\bibitem{chiribella2010probabilistic}
G.~Chiribella, G.~M. D'Ariano, and P.~Perinotti, ``Probabilistic theories with
  purification,'' {\em Physical Review A}, vol.~81, no.~6, p.~062348, 2010.

\bibitem{chiribella2011informational}
G.~Chiribella, G.~M. D'Ariano, and P.~Perinotti, ``Informational derivation of
  quantum theory,'' {\em Physical Review A}, vol.~84, no.~1, p.~012311, 2011.

\bibitem{barrett2007information}
J.~Barrett, ``Information processing in generalized probabilistic theories,''
  {\em Physical Review A}, vol.~75, no.~3, p.~032304, 2007.

\bibitem{barnum2017ruling}
H.~Barnum, C.~M. Lee, C.~M. Scandolo, and J.~H. Selby, ``Ruling out
  higher-order interference from purity principles,'' {\em Entropy}, vol.~19,
  no.~6, p.~253, 2017.

\bibitem{lee2017no}
C.~M. Lee and J.~H. Selby, ``A no-go theorem for theories that decohere to
  quantum mechanics,'' {\em arXiv preprint arXiv:1701.07449}, 2017.

\bibitem{lee2018beyond}
C.~Lee, ``Beyond quantum,'' {\em New Scientist}, vol.~238, no.~3182,
  pp.~28--29, 2018.

\bibitem{galley2016classification}
T.~D. Galley and L.~Masanes, ``Classification of all alternatives to the born
  rule in terms of informational properties,'' {\em arXiv preprint
  arXiv:1610.04859}, 2016.

\bibitem{Fermionic1}
G.~M. D'Ariano, F.~Manessi, P.~Perinotti, and A.~Tosini, ``Fermionic
  computation is non-local tomographic and violates monogamy of entanglement,''
  {\em Europhys. Lett.}, vol.~107, no.~2, p.~20009, 2014.

\bibitem{Fermionic2}
G.~M. D'Ariano, F.~Manessi, P.~Perinotti, and A.~Tosini, ``The {F}eynman
  problem and fermionic entanglement: Fermionic theory versus qubit theory,''
  {\em Int. J. Mod. Phys. A}, vol.~29, no.~17, p.~1430025, 2014.

\bibitem{Wootters-real}
W.~K. Wootters, ``Local accessibility of quantum states,'' in {\em Complexity,
  entropy and the physics of information} (W.~H. Zurek, ed.), pp.~39--46,
  Westview Press, 1990.

\bibitem{d'ariano2014determinism}
G.~D'Ariano, F.~Manessi, and P.~Perinotti, ``Determinism without causality,''
  {\em Physica Scripta}, vol.~2014, no.~T163, p.~014013, 2014.

\bibitem{massar2015hyperdense}
S.~Massar, S.~Pironio, and D.~Pital{\'u}a-Garc{\'\i}a, ``Hyperdense coding and
  superadditivity of classical capacities in hypersphere theories,'' {\em New
  Journal of Physics}, vol.~17, no.~11, p.~113002, 2015.

\bibitem{janotta2011limits}
P.~Janotta, C.~Gogolin, J.~Barrett, and N.~Brunner, ``Limits on nonlocal
  correlations from the structure of the local state space,'' {\em New Journal
  of Physics}, vol.~13, no.~6, p.~063024, 2011.

\bibitem{spekkens2007evidence}
R.~W. Spekkens, ``Evidence for the epistemic view of quantum states: A toy
  theory,'' {\em Physical Review A}, vol.~75, no.~3, p.~032110, 2007.

\bibitem{janotta2013generalized}
P.~Janotta and R.~Lal, ``Generalized probabilistic theories without the
  no-restriction hypothesis,'' {\em Physical Review A}, vol.~87, no.~5,
  p.~052131, 2013.

\bibitem{popescu1998causality}
S.~Popescu and D.~Rohrlich, ``Causality and nonlocality as axioms for quantum
  mechanics,'' in {\em Causality and Locality in Modern Physics}, pp.~383--389,
  Springer, 1998.

\bibitem{popescu-review}
S.~Popescu, ``Non-locality beyond quantum mechanics,'' {\em Nature Physics 10,
  264-270}, 2014.

\bibitem{cavalcanti2021witworld}
P.~J. Cavalcanti, J.~H. Selby, J.~Sikora, T.~D. Galley, and A.~B. Sainz,
  ``Witworld: A generalised probabilistic theory featuring post-quantum
  steering,'' {\em arXiv preprint arXiv:2102.06581}, 2021.

\bibitem{lee2016information}
C.~M. Lee and M.~J. Hoban, ``The information content of systems in general
  physical theories,'' {\em arXiv preprint arXiv:1606.06801}, 2016.

\bibitem{lee2016generalised}
C.~M. Lee and J.~H. Selby, ``Generalised phase kick-back: the structure of
  computational algorithms from physical principles,'' {\em New Journal of
  Physics}, vol.~18, no.~3, p.~033023, 2016.

\bibitem{barnum2014higher}
H.~Barnum, P.~Markus, C.~Ududec, {\em et~al.}, ``Higher-order interference and
  single-system postulates characterizing quantum theory,'' {\em New Journal of
  Physics}, vol.~16, no.~12, p.~123029, 2014.

\bibitem{lee2015computation}
C.~M. Lee and J.~Barrett, ``Computation in generalised probabilisitic
  theories,'' {\em New Journal of Physics}, vol.~17, no.~8, p.~083001, 2015.

\bibitem{landscape}
J.~Barrett, N.~de~Beaudrap, M.~J. Hoban, and C.~M. Lee, ``The computational
  landscape of general physical theories,'' {\em npj Quantum Information},
  vol.~5, no.~1, pp.~1--10, 2019.

\bibitem{lee2015proofs}
C.~M. Lee and M.~J. Hoban, ``Bounds on the power of proofs and advice in
  general physical theories,'' {\em Proc. R. Soc. A 472 (2190), 20160076},
  2016.

\bibitem{lee2016deriving}
C.~M. Lee and J.~H. Selby, ``Deriving grover's lower bound from simple physical
  principles,'' {\em New Journal of Physics}, vol.~18, no.~9, p.~093047, 2016.

\bibitem{lee2016higher}
C.~M. Lee and J.~H. Selby, ``Higher-order interference in extensions of quantum
  theory,'' {\em Foundations of Physics, Volume 47, Issue 1, pp 89--112}, 2017.

\bibitem{lee2017oracles}
C.~M. Lee, J.~H. Selby, and H.~Barnum, ``Oracles and query lower bounds in
  generalised probabilistic theories,'' {\em arXiv:1704.05043}, 2017.

\bibitem{abramsky2004categorical}
S.~Abramsky and B.~Coecke, ``A categorical semantics of quantum protocols,'' in
  {\em Logic in Computer Science, 2004. Proceedings of the 19th Annual IEEE
  Symposium on}, pp.~415--425, IEEE, 2004.

\bibitem{coecke2016picturing}
B.~Coecke and A.~Kissinger, ``Picturing quantum processes. a first course in
  quantum theory and diagrammatic reasoning,'' {\em Cambridge University
  Press}, 2016.

\bibitem{selby2020compositional}
J.~H. Selby and C.~M. Lee, ``Compositional resource theories of coherence,''
  {\em Quantum}, vol.~4, p.~319, 2020.

\bibitem{lee2016thesis}
C.~M. Lee, ``Bounds on computation from physical principles,'' {\em DPhil
  thesis, University of Oxford}, 2016.

\bibitem{prugovevcki1977information}
E.~Prugove{\v{c}}ki, ``Information-theoretical aspects of quantum
  measurement,'' {\em International Journal of Theoretical Physics}, vol.~16,
  no.~5, pp.~321--331, 1977.

\bibitem{busch1989determination}
P.~Busch and P.~J. Lahti, ``The determination of the past and the future of a
  physical system in quantum mechanics,'' {\em Foundations of Physics},
  vol.~19, no.~6, pp.~633--678, 1989.

\bibitem{masanes2013existence}
L.~Masanes, M.~P. M{\"u}ller, R.~Augusiak, and D.~P{\'e}rez-Garc{\'\i}a,
  ``Existence of an information unit as a postulate of quantum theory,'' {\em
  Proceedings of the National Academy of Sciences}, vol.~110, no.~41,
  pp.~16373--16377, 2013.

\bibitem{flammia2005minimal}
S.~T. Flammia, A.~Silberfarb, and C.~M. Caves, ``Minimal informationally
  complete measurements for pure states,'' {\em Foundations of Physics},
  vol.~35, no.~12, pp.~1985--2006, 2005.

\bibitem{coecke2014terminality}
B.~Coecke, ``Terminality implies non-signalling,'' {\em arXiv preprint
  arXiv:1405.3681}, 2014.

\bibitem{hardy2012limited}
L.~Hardy and W.~K. Wootters, ``Limited holism and real-vector-space quantum
  theory,'' {\em Foundations of Physics}, vol.~42, no.~3, pp.~454--473, 2012.

\bibitem{morimae2015power}
T.~Morimae and H.~Nishimura, ``Power of quantum computing with restricted
  postselections,'' {\em arXiv preprint arXiv:1502.00067}, 2015.

\bibitem{morimae2016modified}
T.~Morimae, H.~Nishimura, and F.~L. Gall, ``Modified group non-membership is in
  awpp,'' {\em arXiv preprint arXiv:1602.06073}, 2016.

\bibitem{de2014computation}
N.~de~Beaudrap, ``On computation with'probabilities' modulo k,'' {\em arXiv
  preprint arXiv:1405.7381}, 2014.

\bibitem{bremner2010classical}
M.~J. Bremner, R.~Jozsa, and D.~J. Shepherd, ``Classical simulation of
  commuting quantum computations implies collapse of the polynomial
  hierarchy,'' in {\em Proceedings of the Royal Society of London A:
  Mathematical, Physical and Engineering Sciences}, p.~rspa20100301, The Royal
  Society, 2010.

\bibitem{chaitin1975theory}
G.~J. Chaitin, ``A theory of program size formally identical to information
  theory,'' {\em Journal of the ACM (JACM)}, vol.~22, no.~3, pp.~329--340,
  1975.

\bibitem{fortnow1998complexity}
L.~Fortnow and J.~Rogers, ``Complexity limitations on quantum computation,'' in
  {\em Computational Complexity, 1998. Proceedings. Thirteenth Annual IEEE
  Conference on}, pp.~202--209, IEEE, 1998.

\bibitem{bennett1997strengths}
C.~H. Bennett, E.~Bernstein, G.~Brassard, and U.~Vazirani, ``Strengths and
  weaknesses of quantum computing,'' {\em SIAM journal on Computing}, vol.~26,
  no.~5, pp.~1510--1523, 1997.

\bibitem{grover1997quantum}
L.~K. Grover, ``Quantum mechanics helps in searching for a needle in a
  haystack,'' {\em Physical review letters}, vol.~79, no.~2, p.~325, 1997.

\bibitem{nielsen2010quantum}
M.~A. Nielsen and I.~L. Chuang, {\em Quantum computation and quantum
  information}.
\newblock Cambridge university press, 2010.

\bibitem{meyer2011uselessness}
D.~A. Meyer and J.~Pommersheim, ``On the uselessness of quantum queries,'' {\em
  Theoretical Computer Science}, vol.~412, no.~51, pp.~7068--7074, 2011.

\bibitem{sorkin1994quantum}
R.~D. Sorkin, ``Quantum mechanics as quantum measure theory,'' {\em Modern
  Physics Letters A}, vol.~9, no.~33, pp.~3119--3127, 1994.

\bibitem{barnum2015entropy}
H.~Barnum, J.~Barrett, M.~Krumm, and M.~P. M{\"u}ller, ``Entropy, majorization
  and thermodynamics in general probabilistic theories,'' {\em arXiv preprint
  arXiv:1508.03107}, 2015.

\bibitem{niestegge2016quantum}
G.~Niestegge, ``Quantum teleportation and grover's algorithm without the
  wavefunction,'' {\em arXiv preprint arXiv:1611.02926}, 2016.

\bibitem{dakic2014density}
B.~Daki{\'c}, T.~Paterek, and {\v{C}}.~Brukner, ``Density cubes and
  higher-order interference theories,'' {\em New Journal of Physics}, vol.~16,
  no.~2, p.~023028, 2014.

\bibitem{gogioso2018density}
S.~Gogioso and C.~M. Scandolo, ``Density hypercubes, higher order interference
  and hyper-decoherence: a categorical approach,'' in {\em International
  Symposium on Quantum Interaction}, pp.~141--160, Springer, 2018.

\bibitem{hefford2020hyper}
J.~Hefford and S.~Gogioso, ``Hyper-decoherence in density hypercubes,'' {\em
  arXiv preprint arXiv:2003.08318}, 2020.

\bibitem{sorkin1995quantum}
R.~D. Sorkin, ``Quantum measure theory and its interpretation,'' {\em arXiv
  preprint gr-qc/9507057}, 1995.

\bibitem{sinha2015superposition}
A.~Sinha, A.~H. Vijay, and U.~Sinha, ``On the superposition principle in
  interference experiments,'' {\em Scientific reports}, vol.~5, 2015.

\bibitem{marshall1979inequalities}
A.~W. Marshall, I.~Olkin, and B.~C. Arnold, {\em Inequalities: theory of
  majorization and its applications}, vol.~143.
\newblock Springer.

\bibitem{gross2010all}
D.~Gross, M.~M{\"u}ller, R.~Colbeck, and O.~C. Dahlsten, ``All reversible
  dynamics in maximally nonlocal theories are trivial,'' {\em Physical review
  letters}, vol.~104, no.~8, p.~080402, 2010.

\bibitem{krumm2019quantum}
M.~Krumm and M.~P. M{\"u}ller, ``Quantum computation is the unique reversible
  circuit model for which bits are balls,'' {\em npj Quantum Information},
  vol.~5, no.~1, pp.~1--8, 2019.

\bibitem{lee2019heart}
C.~Lee, ``Heart of quantum,'' {\em New Scientist}, vol.~241, no.~3221,
  pp.~38--41, 2019.

\bibitem{pawlowski2009information}
M.~Pawlowski, T.~Paterek, D.~Kaszlikowski, V.~Scarani, A.~Winter, and
  M.~Zukowski, ``Information causality as a physical principle,'' {\em arXiv
  preprint arXiv:0905.2292}, 2009.

\bibitem{fritz2013local}
T.~Fritz, A.~Sainz, R.~Augusiak, J.~B. Brask, R.~Chaves, A.~Leverrier, and
  A.~Ac{\'\i}n, ``Local orthogonality as a multipartite principle for quantum
  correlations,'' {\em Nature Communications}, vol.~4, 2013.

\bibitem{barrett2005no}
J.~Barrett, L.~Hardy, and A.~Kent, ``No signaling and quantum key
  distribution,'' {\em Physical review letters}, vol.~95, no.~1, p.~010503,
  2005.

\bibitem{blind}
A.~Broadbent, J.~Fitzsimons, and E.~Kashefi, ``Universal blind quantum
  computation,'' in {\em Foundations of Computer Science, 2009. FOCS'09. 50th
  Annual IEEE Symposium on}, pp.~517--526, IEEE, 2009.

\bibitem{blind3}
T.~Morimae and K.~Fujii, ``Blind quantum computation protocol in which alice
  only makes measurements,'' {\em Physical Review A}, vol.~87, no.~5,
  p.~050301, 2013.

\end{thebibliography}

\end{document}